\long\def\symbolfootnote[#1]#2{\begingroup%
\def\thefootnote{\fnsymbol{footnote}}\footnote[#1]{#2}\endgroup}
\def\aj{AJ}
\def\araa{ARA\&A}
\def\apj{ApJ}
\def\aap{A\&A}
\def\aapr{A\&A~Rev.}
\def\mnras{MNRAS}

\newcommand{\msun}{\textup{M}_{\odot}}
\newcommand{\rsun}{\textup{R}_{\odot}}

\newcommand{\msunpc}{\textup{M}_{\odot}\,{\textup{pc}}^{-3}}

\newcommand{\kms}{\textup{km}\,{\textup{s}}^{-1}}

\newcommand{\rhoh}{\rho_{\textup{\scriptsize{h}}}}

\newcommand{\tcr}{t_{\textup{\scriptsize{cr}}}}

\newcommand{\reff}{r_{\textup{\scriptsize{eff}}}}

\newcommand{\dr}{\textup{d}}

\newcommand{\mdyn}{M_{\textup{\scriptsize{dyn}}}}

\newcommand{\mdynobs}{M_{\textup{\scriptsize{dyn}}}^{\textup{\scriptsize{obs}}}}
\newcommand{\mphot}{M_{\textup{\scriptsize{phot}}}}
\newcommand{\sigobs}{\sigma_{\textup{\scriptsize{obs}}}}
\newcommand{\sigdyn}{\sigma_{\textup{\scriptsize{dyn}}}}
\newcommand{\vorb}{v_{\textup{\scriptsize{orb}}}}

\newcommand{\voned}{v_{\textup{\scriptsize{1D}}}}
\newcommand{\sigbin}{\sigma_{\textup{\scriptsize{bin}}}}

\newcommand{\hs}{\hspace{-0.06cm}}
\newcommand{\hsss}{\hspace{-0.03cm}}
\newcommand{\logage}{\log(\textup{age/yr})}
\newcommand{\fbin}{{\rm PDF}(\sigbin^2)}

\newcommand{\frsg}{f_\textup{\scriptsize{SG}}}
\newcommand{\pcrit}{P_\textup{\scriptsize{crit}}}

\newcommand{\siglev}{\textup{s.l.}}
\newcommand{\nrsg}{200}
\newcommand{\nsimu}{1000}
\newcommand{\np}{N_\textup{\scriptsize{p}}}

\def\apgt{\ {\raise-.5ex\hbox{$\buildrel>\over\sim$}}\ }
\def\aplt{\ {\raise-.5ex\hbox{$\buildrel<\over\sim$}}\ }

\documentclass[useAMS,usenatbib, times]{mn2e}
\usepackage{amssymb,latexsym,graphicx,natbib,eufrak,times}

\title[On the velocity dispersion of young star clusters]
{On the velocity dispersion of young star clusters: super-virial or binaries?}
\author[M. Gieles et al.]
{M.~Gieles$^{1}$\footnote{}, H.~Sana$^{1,2}$ and S.~F.~Portegies~Zwart$^3$\\
$^1$ European Southern Observatory, Casilla 19001, Santiago 19, Chile \\
$^2$ Astronomical Institute `Anton Pannekoek', University of Amsterdam,
Kruislaan 403, 1098 SJ Amsterdam, the Netherlands\\
$^3$ Leiden Observatory, Leiden University, PO Box 9513, 2300 RA Leiden, the Netherlands
}
\date{Accepted 2009 November 5. Received 2009 October 26; in original form 2009 September 23}

\def\LaTeX{L\kern-.36em\raise.3ex\hbox{a}\kern-.15em
 T\kern-.1667em\lower.7ex\hbox{E}\kern-.125emX}

\begin{document}         
\maketitle
\begin{abstract}
Many young extra-galactic clusters have a measured velocity dispersion
that is too high for the mass derived from their age and total
luminosity, which has led to the suggestion that they are not in
virial equilibrium. Most of these clusters are confined to a narrow
age range centred around 10\,Myr because of observational constraints.
At this age the cluster light is dominated by luminous evolved stars,
such as red supergiants, with initial masses of $\sim$13$-22\,\msun$
for which (primordial) binarity is high.  In this study we investigate
to what extent the observed excess velocity dispersion is the result
of the orbital motions of binaries.  We demonstrate that estimates for
the dynamical mass of young star clusters, derived from the observed
velocity dispersion, exceed the photometric mass by up-to a factor of
10  and are consistent with a constant offset in the square of the
velocity dispersion.  This can be reproduced by models of virialised
star clusters hosting a massive star population of which $\sim$25\% is
in binaries, with typical mass ratios of $\sim$0.6 and periods of
$\sim$1000 days. We conclude that binaries play a pivotal role in
deriving the dynamical masses of young ($\sim$10\,Myr) moderately
massive and compact ($\lesssim10^5\,\msun; \gtrsim 1\,$pc) star
clusters.
\end{abstract}


\section{Introduction}

Young massive clusters have received considerable attention in the
last decade because they trace star formation
(e.g. \citealt{1995AJ....109..960W, 1997AJ....114.2381M,
1999AJ....118..752Z}). Advances in observations enabled us to resolve
such star clusters up to $\sim 20$\,Mpc, allowing determination of
their fundamental parameters, such as mass and radius
(e.g. \citealt{2004A&A...416..537L}).

The mass of a resolved star cluster can be determined in two ways: one
of them by converting the observed luminosity, age and distance
directly to mass via the age dependent mass-to-light ratio ($M/L$)
taken from a single stellar population (SSP) model.  We refer to the
resulting mass as the photometric mass $\mphot$. This method requires
an estimate of the cluster age, which again requires estimates for the
metallicity and the stellar initial mass function (IMF).

An independent mass estimate is based on the virial theorem and this
mass is generally referred to as the dynamical
mass \citep{1987degc.book.....S}:

\begin{equation}
\mdyn=\frac{\eta\sigdyn^2\reff}{G}.
\label{eq:mdyn}
\end{equation}
Here $G$ is the gravitational constant, $\sigdyn$ is the line of sight
velocity dispersion in the cluster, $\reff$ is the effective
(half-light) radius\footnote{Here we assume that the half-light radius
is the same as the half-mass radius, which is not the case when the
cluster is mass
segregated \citep[][]{2006MNRAS.369.1392F,2008MNRAS.391..190G}.}
and $\eta\simeq9.75$ is a constant that depends slightly on the
density profile.

Equation~(\ref{eq:mdyn}) is valid for a cluster in virial equilibrium
consisting of single stars. Since in this study we consider possible
difference between $\sigdyn$ and the observed velocity dispersion,
$\sigobs$, we will refer to the empirically derived dynamical mass,
i.e. based on $\sigobs$, as $\mdynobs$.

A comparison between $\mphot$ and $\mdynobs$ serves as a check for the
range of assumptions on which both mass estimates are based.  An
inconsistency between $\mphot$ and $\mdynobs$ can be attributed to
variations in the IMF, on which $\mphot$ is in part based, or to a
lack of virial equilibrium, on which $\mdynobs$ is based.  For many
young ($\sim$10\,Myr) star clusters $\mdynobs>\mphot$, with $\mdynobs$
up to $\sim$10 times larger than $\mphot$ \citep[e.g.][hereafter
B06]{2006A&A...448..881B}, suggesting that these objects are
super-virial. For older clusters ($ \gtrsim100\,$Myr) there is good
agreement between $\mdynobs$ and $\mphot$
\citep[e.g.][B06]{2004AJ....128.2295L}.

The alleged super-virial state of some young clusters has been
attributed to the impulsive expulsion of residual gas from the parent
molecular cloud in which the star cluster
formed \citep[e.g.][]{2006MNRAS.373..752G}.  Such early outgassing,
driven by stellar winds of massive stars or supernovae, causes the
stellar velocities to be high compared to the binding energy of the
stars. This argument has been used to motivate infant mortality of
young star clusters \citep{2003ARA&A..41...57L}.

However, the gas expulsion theory has difficulties in explaining
the super-virial velocity for the 10\,Myr old clusters presented in
B06. The arguments are as follows: The time needed to completely
dissolve, or to find a new virial equilibrium after impulsive gas
expulsion is about 20 crossing times, $\tcr$, where
$\tcr\propto\rhoh^{-1/2}$ and $\rhoh$ is the density within the
half-mass radius \citep[see for example Fig.~8
in][]{2007MNRAS.380.1589B}.  Hence, to be able to `catch' an unbound
or expanding cluster at 10 Myr, $\tcr$ should be $\gtrsim1\,$Myr. This
corresponds to a half-mass density of stars and gas of
$\rhoh\lesssim300\,\msunpc$. Clusters with shorter $\tcr$ (higher
density) have expanded into the field, or found a new equilibrium, a
few Myrs after gas expulsion and are not observable as super-virial
clusters at 10\,Myr\footnote{The models of \citet{2006MNRAS.373..752G}
start with a density of $\sim60\,\msunpc$ ($\tcr\approx2.5\,$Myr) in
the embedded phase, and this is why they find that the effects of gas
expulsion are observable for 25 Myr.}.  The density in the embedded
phase of the clusters under discussion is unknown, but can be roughly
estimated using their current densities. The present day densities are
$\rhoh\approx10^{3\pm1}\,\msunpc$ (Table~\ref{tab:data}). The
densities in the embedded phase were at least a factor $1/\epsilon^4$
higher, where $\epsilon$ is the star formation efficiency.  This
because the mass of the embedded cluster has reduced by a factor
$\epsilon$ and the cluster has expanded at least by a factor of
$1/\epsilon$ as a response to it, contributing a factor $1/\epsilon^3$
to the reduction of $\rhoh$. The $1/\epsilon$ expansion holds for
adiabatic mass loss, for impulsive mass loss and $\epsilon\lesssim0.9$
the cluster expands much more \citep{1980ApJ...235..986H}. So for the
clusters at 10\,Myr the estimated densities in the embedded phase are
much too high to still have features of gas expulsion detectable in
their velocity dispersion at 10\,Myr. These arguments suggest that
deviations from virial equilibrium are not a plausible explanation and
an alternative explanation for the high $\sigobs$ values is needed.  

The existence of binary stars is generally ignored in the estimates
for $\mdynobs$, even though their internal velocities can lead to an
over estimation of $\mdyn$ (Kouwenhoven \& de Grijs 2008, hereafter
K08)\nocite{2008A&A...480..103K}.  K08 studied this phenomenon in
virialised star clusters with a 100\%\ binary fraction and a range of
$\sigdyn$. They subsequently derive $\mdynobs$ by `measuring'
$\sigobs$ and applying equation~(\ref{eq:mdyn}).  They found that the
presence of binaries can lead to an overestimation of $\mdyn$ by a
factor of $\sim$2 for clusters with $\sigdyn\simeq1\,\kms$.  For
clusters with $\sigdyn\simeq10\,\kms$ they found only a 5\% increase
in $\mdynobs$ due to binaries. They therefore concluded that binaries
are not important for massive/dense clusters.  \citet[][hereafter
M08]{2008A&A...489.1091M} found $\mdynobs/\mphot \simeq 10$ for some
of the star clusters in the Antennae galaxies (NGC~4038/4039) and
NGC~1487 and since these clusters have velocity dispersions of
$10-20\,\kms$ they subsequently concluded that binaries are not
important and that these star clusters are super-virial and dissolving
quickly.

Here we revisit the effect of binaries on $\mdynobs/\mphot$ and we
focus on $\sim$10\,Myr old star clusters.  This is motivated by our
desire to incorporate the effect of the steep increase of the stellar
luminosity with increasing stellar mass, which is in particular
important for young clusters, an aspect not considered by K08.  Of the
two approaches presented by K08, one focused on solar-type stars, the
other uses a Kroupa IMF for the primary stars. They give equal weight
to each binary in their computed velocity dispersion.  In addition K08
do not consider stars more massive than $20\,\msun$. This approach may
be appropriate for studies of intermediate age and old open star
clusters, but it is less suitable for young star clusters.

At an age of $\sim$10\,Myr the cluster light is dominated by the most
massive ($\gtrsim 15\,\msun$) stars for which binarity is high and
ignoring them can lead to misinterpretations of observations of
various astrophysical processes \citep[e.g.][]{1998A&ARv...9...63V}.
Massive binaries have a larger effect on $\sigobs$ than low-mass
binaries due to their higher orbital velocities, but also due to the
more common short-periods and comparable
masses \citep[e.g.][]{1991A&A...248..485D, 2002ApJ...574..762P,
2008MNRAS.386..447S, SGE2009,2009AJ....137.3358M}.  Incorporating the
massive stars in our calculation has two important effects, both of
which amplify the effect of binarity on $\sigobs$ with respect to the
results of K08: massive stars dominate the cluster light and their
higher masses and (intrinsically) different binary properties give
rise to a larger $\sigobs$.

In this paper we quantify the effect of the presence of (massive)
binaries on $\mdynobs/\mphot$ and we use this ratio is a proxy of the excess
dispersion. In Section~\ref{sec:model} we discuss the properties of
the binary population that is expected in young ($\sim$10\,Myr)
clusters and we present a simple model for the additional velocity
dispersion due to such binaries.  In Section~\ref{sec:data} we
summarize existing observational results to confront our model with.
Our conclusions are discussed in Section~\ref{sec:conc}. All the
specific acronyms used in this study and their definitions are given
in Table~\ref{tab:acronyms}.

\section{The velocity dispersion due to binarity}
\label{sec:model}

\subsection{The importance of massive binaries}
\label{ssec:obs_bin}
The young clusters with measured $\mdynobs$ and $\mphot$ have a rather
narrow range in ages of $\sim$$8-13$\,Myr.  This is mainly because of
the onset of red supergiants in this age range making clusters
brighter and easier to detect and study in detail.  Stars in a
stellar population with an age of 10\,Myr have initial mass of
$13-22\,\msun$, corresponding to masses of $13-16\,\msun$\ at an age
of 10\,Myr \citep{2001A&A...366..538L}.   If we would consider a
small spread around 10\,Myr the quoted mass range would be slightly
larger, but for simplicity will continue with the assumption of a
constant age of 10\,Myr.  Those massive stars appear to have high
primordial multiplicity with a spectroscopic binary fraction of
$\sim$50\% or more (i.e.
$f\gtrsim0.5$, \citealt{2009AJ....137.3358M,2009AJ....137.3437B}).

Most of the measurements we discuss in Section~\ref{sec:data} are done
in the near infrared. At these wavelengths red supergiants (RSGs)
dominate the observed light and therewith the measured $\sigobs$. For
the studies done in the optical wavelength the blue supergaints (BSGs)
are more dominant. We here refer to the population of luminous 
evolved stars as supergiants (SGs) and use the subscripts SG to denote
parameters that apply to these stars.

Since the SGs outshine the main sequence stars it is important to
establish the binary fraction among them.  This estimate is
complicated by the internal evolution of binary stars affecting
especially the RSG phase and hence the actual population of SGs
present at an age of 10\,Myr. In particular, short-period binaries are
likely to experience a common envelope evolution (CEE) and/or
Roche-lobe overflow (RLOF) which causes the binary components to
follow a different evolution compared to single stars of similar
initial mass, and may prevent the RSG stage altogether.

\citet[][hereafter E08]{2008MNRAS.384.1109E} 
find that these effects reduce the average duration of the RSG phase
by a factor of two or three.  They find this for a population of
binaries with a flat distribution in $\log P/\dr$ between $-0.15$ and
$4.5$ and a flat distribution of $q$ between 0.1 and 0.9,  where
$P$ and $q$ are the orbital period and the ratio of the secondary mass
over the primary mass, respectively.  The short period binaries with
high mass ratios are most affected by interactions through RLOF and
CEE. For our simple model we assume a minimum period, $\pcrit$, and as
an approximation of the shortened evolved phase of primaries in tight
binaries we remove the binaries with $P<\pcrit$. The fraction of
binaries we remove should roughly match the fractional reduction of
the average life-time of the RSGs (factor of $2-3$). This constraint
is met for $\pcrit=500\,\dr$ since $63\%$ of the binaries in the E08
population have $P<500\,\dr$ for $m_1=15\,\msun$.
  
RSGs at 10\,Myr have a maximum radius of $\sim$900$\,\rsun$. For
$m_1=15\,\msun$ and $q=0.6$ this corresponds roughly to the separation
of a binary with $\pcrit$. For $P=2000\,\dr$ the Roche-lobe radius is
around $900\,\rsun$ \citep[using the formula
of][]{1983ApJ...268..368E} and binaries with longer orbital periods
will follow an evolutionary path similar to single stars (08).  
So in our model we remove all binaries with $P<500\,\dr$ and assume
that binaries with $P>500\,\dr$ experience a SG phase unaffected by
binary evolution, even though it is expected that the RSG phase of
primaries in binaries with $500<P/\dr<2000$ is affected by the
companion. In reality it will not be such a step function, since most
SGs do contribute at some stage in their evolution to the integrated
light. But under our assumptions we reduce the number of binaries
roughly by the same fraction as what was found for the fractional
reduction of the average RSG phase in the model of E08. By removing
all binaries with $P<500\,\dr$ we are probably making a conservative
approach since we bias our binary population to longer periods. In
reality these binaries can continue to contribute to the velocity
dispersion. This because the primary does not necessarily becomes dark
after its shortened RSG phase, and if it does, the secondary can still
contribute to the velocity dispersion (E08). 

The relevant parameter for studying the binaries that contribute to
the velocity dispersion is the fraction of binaries among SGs, which
we identify with $\frsg$. Using $\np$ for the number of stars with
initial masses in the range $13-22\,\msun$ and the fraction of
binaries with an orbital period $P > \pcrit$ as $g$, then the number
of stars in binaries unaffected by interaction is $gf\np$  and the
number of SGs that is removed is $(1-g)f\np$.  The total number of
remaining SGs, i.e. single and in binaries, is $(1-f)\np+gf\np$. So we
can write for $\frsg$
\begin{equation}
     \frsg = \frac{g f} {(1 -f) + g f}.
\label{eq:frsg}
\end{equation}
In equation~(\ref{eq:frsg}) we have neglected the possibility that
secondary stars contribute to the SG population, thus slightly
underestimating $\frsg$.  If all stars are in binaries ($f=1$) then
$\frsg=f$ for all values of $g$.  For the remainder of our analysis we
adopt a more conservative value of $f=0.6$ in our parametric model
(Section~\ref{ssec:model}) and a range $0.3<f<0.9$ for the Monte Carlo
simulations in Section~\ref{ssec:mc}.

The orbital periods of early-type spectroscopic binaries range from a
couple of days to about 10~years. Adopting an {\"O}pik's law in the
interval $0.3<\log P/\dr<3.5$ and a period threshold
$\log\pcrit/\dr=2.7$\ we find $g=0.25$ (i.e. we remove 75\% of the
binaries), which via equation~(\ref{eq:frsg}) results in $\frsg \simeq
0.25$. For the representative period we use $P=10^3\,\dr$, which is
approximately the mean of the periods above $\pcrit$ when assuming a
flat distribution in $\log P$.

The distribution of mass ratios for high-mass stars
appears to be flat between $q \simeq 0.2$ (the typical detection limit
for SB2 systems) and $q=1$ \citep[e.g.][]{IHOT09}. We adopt $q=0.6$ as
a typical value for the mass ratio.

Our adopted values of the parameters that control the SG binary
population at 10\,Myr are summarised in Table~\ref{tab:param}.  These
values serve as input for the model presented in the next section.

\begin{table}
\caption{Overview of the specific  acronyms used in this study.}
\label{tab:acronyms}
\resizebox{\columnwidth}{!}{
\begin{tabular}{ll}
\hline
Acronym & Description\\\hline
$f$             & primordial binary fraction of massive stars ($13-22\,\msun$)\\
$g$             & fraction of primordial binaries unaffected by interaction \\ 
$\frsg$         & effective binary fraction among SGs at 10\,Myr\\
$m_1$           & mass of the primary star\\
$\np$           & number of stars with initial masses in the range $13-22\,\msun$\\
$q$	        & ratio of the secondary mass over the primary mass\\
$P$             & orbital period\\
$\pcrit$        & minimum period for binaries to be unaffected by interaction\\
$\vorb$         & orbital velocity of the primary star\\
$\voned$        & line of sight velocity of the primary star\\\hline
$\reff$         & cluster half-light radius in projection\\
$\sigdyn$       & 1D dynamical velocity dispersion of cluster members\\
$\sigobs$       & empirically determined 1D velocity dispersion\\
$\sigbin$       & 1D velocity dispersion due to binary orbital motions\\
$\mdyn$         & dynamical cluster mass based on $\sigdyn^2$\\
$\mdynobs$      & empirically  determined dynamical mass based on $\sigobs^2$ \\
$\mphot$        & photometric cluster mass\\
\hline
\end{tabular}
}
\end{table}

\subsection{A parametric model for the velocity contribution of binaries}
\label{ssec:model}
To quantify the importance of binaries on $\sigobs$ we model their
observational characteristics.  Since the dynamical velocities of the
cluster members (stars and centres of mass of binaries) and the
orbital velocities of the binary members are uncorrelated, we can
write $\sigobs^2=\sigdyn^2+\sigbin^2$. Here we derive a simple
expression for the contribution to $\sigobs^2$ of the orbital motions
of binaries, $\sigbin^2$.

Since the secondary is generally much fainter than the primary we
ignore its contribution to the light and focus only on the primary
star.  Its orbital velocity, $\vorb$, can be expressed in terms of
$q$, $m_1$ and $P$ using Kepler's third law:
\begin{equation}
     \vorb = q \left( \frac{2}{1+q} \right)^{2/3}
               \left( \frac{\pi G m_1}{2P} \right)^{1/3}.
\label{eq:vorb}
\end{equation}

The contribution to the line of sight velocity, $\voned$, depends on
the inclination, $i$, of the orbital plane and the phase, $\theta$, in
which the binary is observed.  We first assume a population of
binaries with the same $q$, $m_1$ and $P$ and random orientations of
the orbital planes and (un-correlated) random orbital phases. This
results in flat distributions of $-1\le\cos(i)\le+1$ and
$0\le\theta\le2\pi$.  For each individual binary
$\voned=\vorb\sin(i)\cos(\theta)$ so the distribution of $\voned$
values is the joint probability density function of $\sin(i)$ and
$\cos(\theta)$ multiplied by $\vorb$, which is flat between $-\vorb$
and $+\vorb$.  The variance of this distribution is
$\sigbin^2=\vorb^2/3$. In reality there will be a spread in the binary
parameters which will make the line of sight velocity distribution
peaked, with a similar variance. We continue with the assumption of a
population of identical binaries to be able to analytically express
our result in the binary parameters.  In Section~\ref{ssec:mc} we
validate this assumption and quantify the expected spread using Monte
Carlo simulations.

Taking into account that only a fraction $\frsg$
(Section~\ref{ssec:obs_bin}) of the stars that contribute to the
cluster light is part of a binary reduces $\sigbin^2$ by a factor
$\frsg$.  The dependence of $\sigbin^2$ on the binary parameters can
then be expressed as
\begin{equation}
     \sigbin^2= \left(\frac{\frsg}{3}\right)
                \left(\frac{2q^{3/2}}{1+q}\right)^{4/3}
                \left(\frac{\pi Gm_1}{2P}\right)^{2/3}.
\label{eq:sigbinsq}
\end{equation}

For the reference values (Table~\ref{tab:param}) we find that
$\sigbin\simeq6.6\,\kms$, which is equal to $\sigdyn$ for a
(virialised) cluster with $M=10^5\,\msun$ and $\reff=1\,$pc
(equation~\ref{eq:mdyn}). So for such clusters and these binary
parameters $\mdynobs$ overestimates the true mass $M$ by a factor of
two because of binaries. We use these scaling values to write a more
general expressing for the ratio
\begin{equation}
\frac{\sigbin^2}{\sigdyn^2}\hs\simeq\hs \left(\hs\frac{\frsg}{0.25}\hs\right)
                            \left(\frac{q}{0.6}\right)^{\hs\frac{3}{2}}\hsss
                            \left(\hs\frac{m_1}{15\msun}\hs\right)^{\hs\frac{2}{3}}\hs
                            \left(\hss\frac{10^3\hsss \dr}{P}\hss\right)^{\hs\frac{2}{3}}\hs
                            \left(\hs\frac{M/\reff}{10^5\msun\textup{pc}^{\hs-1}}\hs\right)^{\hs-1}
                            \hs\hs\hs\hs\hs\hs,
\label{eq:ratio}
\end{equation}
where we have approximated the term $[2q^{3/2}/({1+q})]^{4/3}$ from
equation~(\ref{eq:sigbinsq}) by $q^{3/2}$. Equation~(\ref{eq:ratio})
is accurate to within 8\% for $q\gtrsim0.2$.

In the next section we will use the ratio $\mdynobs/M$ as a measure of
the excess dispersion, which we can write as
\begin{eqnarray}
\frac{\mdynobs}{M}&=&\frac{\sigdyn^2+\sigbin^2}{\sigdyn^2},\label{eq:mratio0}\\
                  &\simeq&1+\left(\frac{M/\reff}{10^5\msun\textup{pc}^{-1}}\right)^{-1},
\label{eq:mratio}
\end{eqnarray}
where in the last step we have used the reference values of
Table~\ref{tab:param} such that the binary part of
equation~(\ref{eq:ratio}) equals 1.

For $M/\reff < 10^5\,\msun\,\textup{pc}^{-1}$ binaries dominate the
measured velocities and therefore $\mdynobs/M\propto(M/\reff)^{-1}$
(for a constant $\sigbin^2$).  For higher values of $M/\reff$ the
presence of binaries has little effect on the estimated mass and
$\mdynobs/M \simeq 1$.

\begin{table}
\caption{Adopted values for the parameters of SG binaries at 10\,Myr.}
\label{tab:param}
\begin{tabular}{lcrrc}
\hline
                     &Reference    &\multicolumn{3}{c}{Range}\\
parameter            &             & min     & max      & distribution\\\hline
$f$                  & $0.6~~$     & $0.3$    &$0.9$    &flat\\
$g$                  & $0.25$      & $-$      &$-$      &$-$\\
$\frsg$              & $0.25$      & $-$      &$-$      &$-$\\
$m_1/\msun$          & $15~~~~~~~$ & $13~~~$  &$16~~~$  &Salpeter\\
$q$                  & $0.6~~$     & $0.2$    &$1.0$    &flat\\
$\log P/\dr$         & $3.0~~$     & $0.3$    & $3.5$   &flat\\
$\log \pcrit/\dr$    &$2.7~~$      &$2.7$     & $2.7$   &$-$\\
\hline
\end{tabular}
\end{table}

\subsection{A Monte Carlo validation}
\label{ssec:mc}
Up to this point we have assumed populations of equal binaries giving
a flat distribution of $\voned$ values and a fixed value for $\sigbin$ for
each cluster.  First we verify the assumption that the shape of the
$\voned$ distribution resembles a Gaussian when a range of binary
parameters is assumed (Section~\ref{sssec:binaries}). Then we quantify
the expected spread in $\sigbin^2$ values when comparing clusters
(Section~\ref{sssec:spread}).

\subsubsection{The velocity dispersion of a binary population}
\label{sssec:binaries}
\begin{figure}
\center\includegraphics[width=8cm]{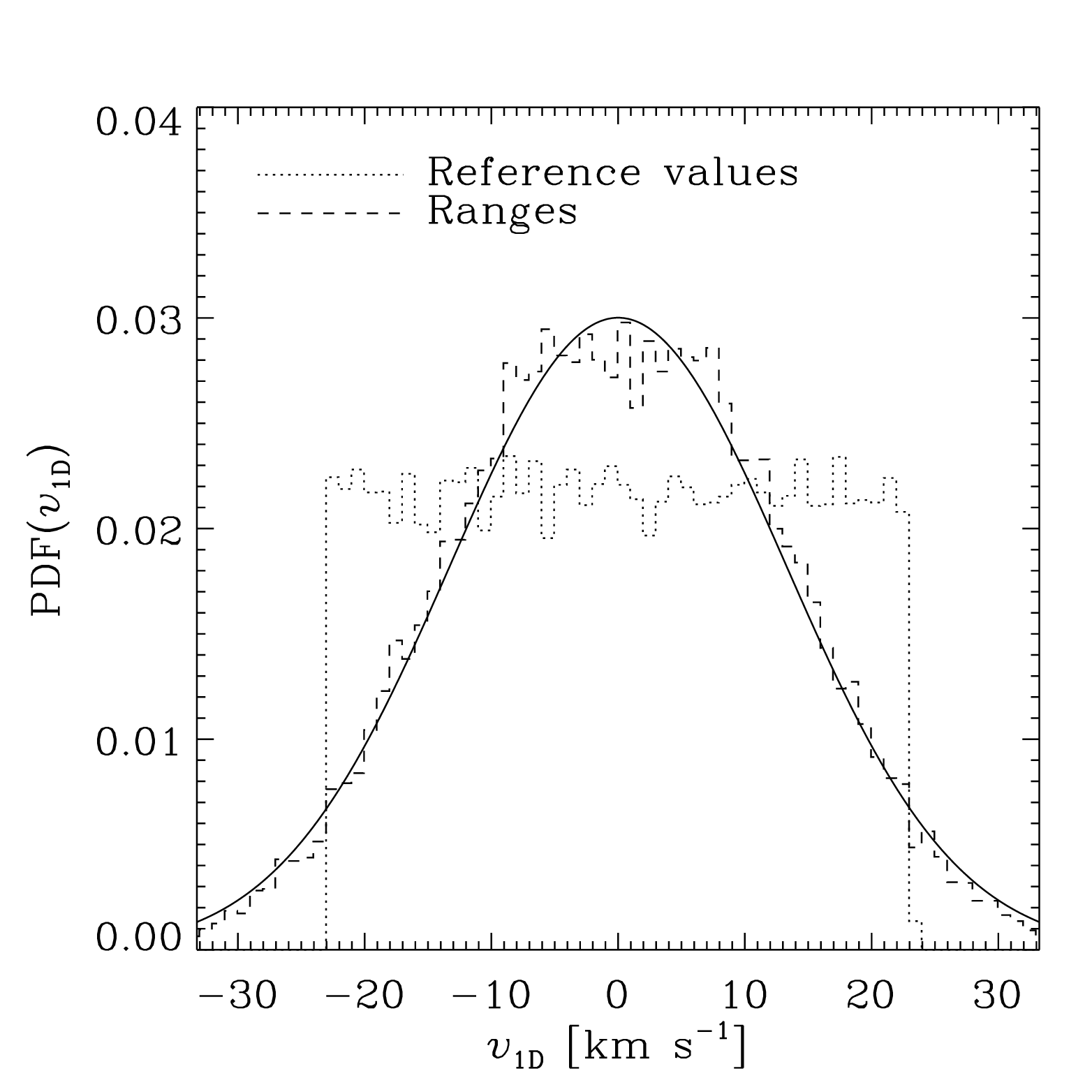}         
    \caption{ Monte Carlo simulations of the probability density
            functions of $\voned$ values of binary populations based
            on the reference values (i.e. all binaries identical,
            dotted histogram) and using a spread in the binary
            parameters (dashed histogram). Both simulations are based
            on the values for $m_1$, $q$ and $\log P$ quoted in
            Table~\ref{tab:param} and both consist of $10^4$ binaries.
            The Gaussian curve shown with a full line is the
            approximation based on the reference values from
            Table~\ref{tab:param}, but using $f=1$
            ($\sigbin=\vorb/\sqrt{3}\approx13\,\kms$). This
            approximation nicely describes the more realistic
            simulation based on a range of values (dashed histogram).
            }  \label{fig:sig}
\end{figure}

We generate two populations of $10^4$ binaries, i.e. no single stars,
to study the shape of the velocity dispersion of their orbital
motions. For one population we give all primaries a $\vorb$ based on
equation~(\ref{eq:vorb}) and the reference values from
Table~\ref{tab:param}. The values of $\voned$ are acquired by
multiplying $\vorb$ for each binary by a random number between $-1$
and $+1$ (Section~\ref{ssec:model}). The resulting distribution and
the Gaussian approximation ($\sigbin=\vorb/\sqrt{3}$,
Section~\ref{ssec:model}) are shown as a  dotted histogram and a
full line, respectively, in Fig.~\ref{fig:sig}. For the second
population we randomly draw values for the masses, mass ratios and
periods from the distributions described in Table~\ref{tab:param}. The
binaries with $P<\pcrit$ are taken out of the sample.  With
equation~(\ref{eq:vorb}) we then calculate $\vorb$ for each remaining
binary and $\voned$ is again acquired by multiplying $\vorb$ by a
random umber between $-1$ and $+1$. The resulting distribution is
shown as a dashed histogram in Fig.~\ref{fig:sig}. Two things can be
seen from this figure: 1.) the width of the more realistic
distribution (i.e. using a range in binary parameters) is well
approximated by our simple model and 2.)  this distribution is close
to Gaussian. This last point is important since we have assumed in
Section~\ref{ssec:model} that we can quadratically add $\sigbin$ to
$\sigdyn$ to get the total velocity dispersion.

\subsubsection{The expected dispersion in the binary dispersion}
\label{sssec:spread}
Here we quantify the spread in $\sigbin^2$, i.e. the dispersion in the
additional velocity dispersion squared, when comparing different
realisations of binary populations, caused by the fact that the number
of binaries is small and that there is a spread in the binary fraction
(Table~\ref{tab:param}).

\begin{figure}
\center\includegraphics[width=8cm]{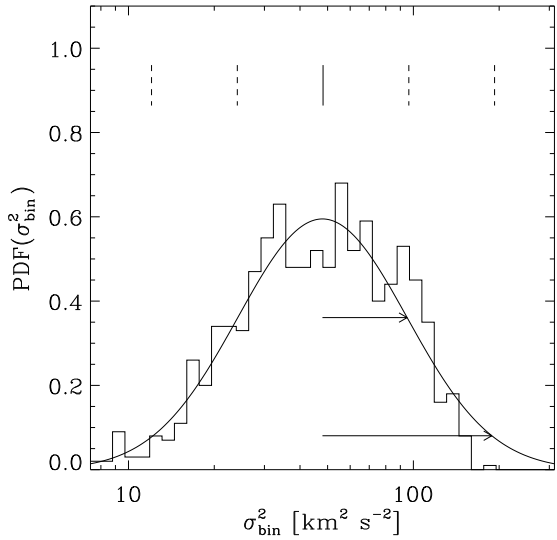}         
    \caption{The probability density function of $\sigbin^2$ values
             following from the Monte Carlo experiment described in
             Section~\ref{ssec:mc} for the adopted distributions in
             $f$, $m_1$, $q$ and $\log P$ (Table~\ref{tab:param}). The
             solid line near the peak of the distribution indicates
             the value of $\sigbin^2$ derived with
             equation~(\ref{eq:sigbinsq}) using the reference values
             from Table~\ref{tab:param}.  The dashed lines indicate
             factors of two of variation. This corresponds
             approximately to the one and two standard deviations
             (horizontal arrows) of the log-normal approximation.
             }  \label{fig:fbin}
\end{figure}

We generate $\nsimu$ massive star populations, each consisting of
$\nrsg$ SGs (an approximate number for a cluster of mass
$10^5\,\msun$, \citealt{2008MNRAS.383..263L}).  For each population,
we randomly sample a value for $f$ and thus have $200\times f$
binaries.  The $\voned$ values of the binaries are calculated in the
same way as in Section~\ref{sssec:binaries} using the ranges from
Table~\ref{tab:param}. For each population the variance of the
1-dimensional velocity distribution ($\sigbin^2$) of the remaining SGs
in the sample (single and binary) is then calculated. The resulting
probability density function (PDF) of the $\sigbin^2$ values is shown
in Fig.~\ref{fig:fbin}. The reference value of $\sigbin^2$ is
indicated with a vertical solid line and is very close to the mode of
the distribution.  When approximating $\fbin$ by a log-normal we find
a standard deviation of $\sim$$0.7$ corresponding to a factor of
$\sim$2 relative to the mode.

We will now compare the model to empirical determinations of
$\mdynobs/M$. Since we do not know the real mass $M$ we use $\mphot$
as a proxy.

\section{Comparison with observations}
\label{sec:data}
We illustrate the effect of the presence of binaries by comparing the
results of our model from the previous section to the empirical ratio
$\mdynobs/\mphot$ for a number of clusters.  The cluster masses
$\mdynobs$ and $\mphot$ follow from literature values for magnitude,
age, $\sigobs$ and $\reff$.  We subsequently re-derive $\mphot$ and
$\mdynobs$ to obtain a homogeneous sample, which is important because
the literature values are derived by a number of groups using a
variety of SSP models to derive $\mphot$ and apply different (small)
corrections to the value of $\eta$ (equation~\ref{eq:mdyn}) because of
mass segregation \citep{2006MNRAS.369.1392F} and variations in the
density profiles (for those clusters for which measurements of their
surface brightness profile are available). All cluster parameters and
references to the relevant literature are given in
Table~\ref{tab:data}.

We use the \citet{2003MNRAS.344.1000B} SSP models with a Chabrier IMF
and solar metallicity to derive $\mphot$. For each cluster the age
dependent $M/L$ is found from the observed cluster age. Combining
$M/L$ with the absolute magnitudes ($M_V$ for 7 clusters, and $M_K$
for the rest) we determine $\mphot$.  The quoted upper and lower
limits in $\mphot$ are calculated through the uncertainties in
$\logage$.  We use equation~(\ref{eq:mdyn}) to determine $\mdynobs$,
with $\eta=9.75$, and the uncertainty is calculated using the
uncertainties in $\sigobs$ and $\reff$ by adopting standard error
propagation. The calculated values for $\mphot$ and $\mdynobs$ are
presented in Table~\ref{tab:data}.

We now assume that our choice for the IMF and the metallicity is
representative for all clusters and that variations in $\eta$ due to
mass segregation and the density profile are negligible.  Under these
assumptions $\mphot$ reflects the true mass $M$ and subsequently
$\sigdyn^2$ scales with $\mphot/\reff$ (equation~{\ref{eq:mdyn}).
 However, $\mphot$ is also affected by binarity since the
shortened RSG phase in short period binaries reduces the integrated
luminosity \citep[recently noted by][]{2009ApJ...696.2014D}. This
effect reduces the fraction of bright stars visible at 10\,Myr by a
factor of $(1-g)f\approx0.5$ (Section~\ref{ssec:obs_bin}).

In Fig.~\ref{fig:data} we present the data.  The trend that clusters
with a small $\mphot/\reff$ tend to have high $\mdynobs/\mphot$, and
which drops with increasing $\mphot/\reff$ is well reproduced by a
population of binaries among the most massive stars. The dispersion in
the observations around the mean value for our model
(equation~[\ref{eq:mratio}], solid curve in Fig.~\ref{fig:data})
roughly corresponds to the spread following from our Monte Carlo
experiment (dashed lines) when allowing a spread in the binary
parameters, rather than fixed values.

\begin{figure}
\includegraphics[width=8cm]{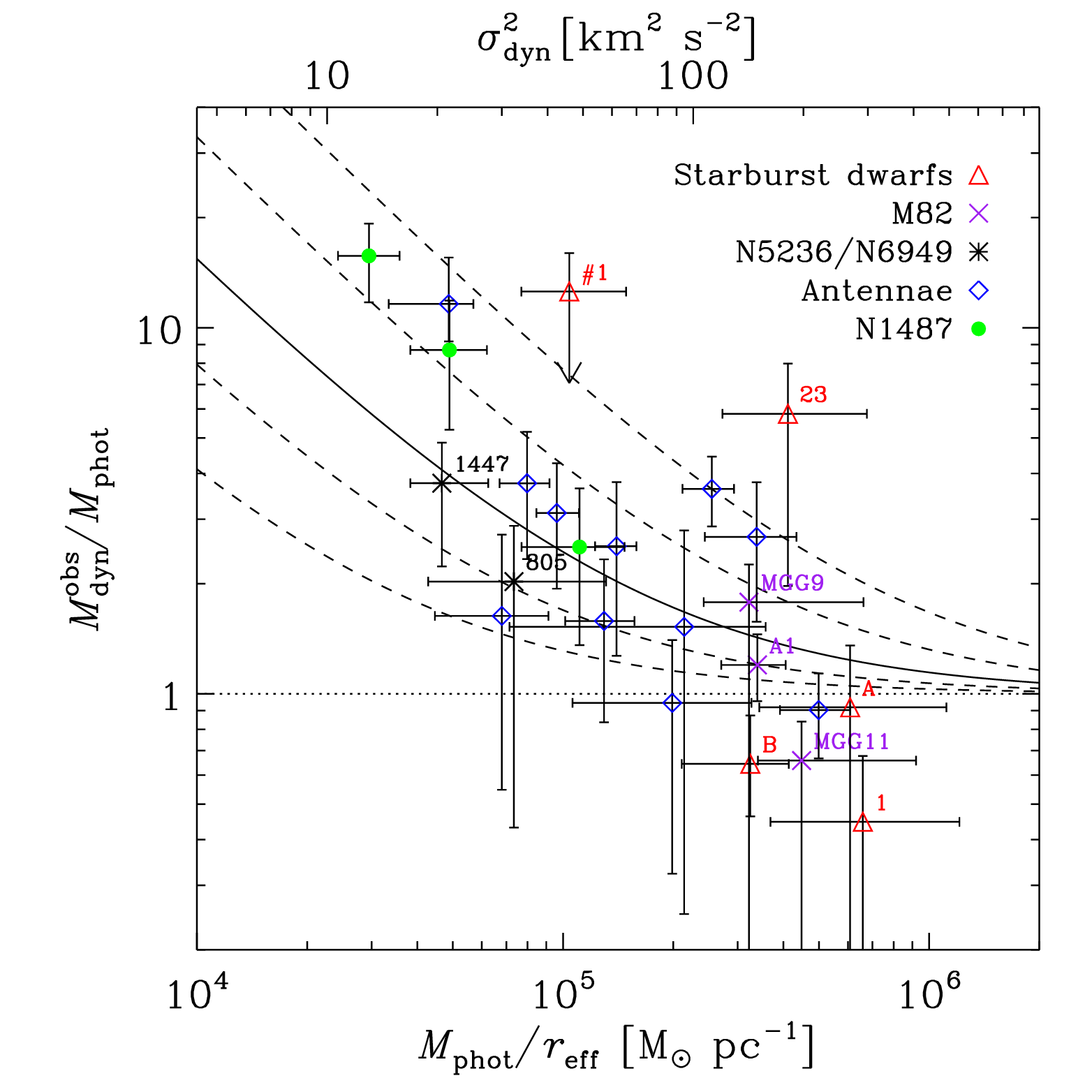}
  \caption{The ratio $\mdynobs/M$ as a function of the square of the
           velocity dispersion expressed in terms of the observables
           $\mphot/\reff$ (equation\,~\ref{eq:mdyn}).  The solid curve
           is calculated assuming the fixed binary parameters adopted
           in Section~\ref{ssec:obs_bin}, which are: $\frsg = 0.25$,
           $q=0.6$, $m_1 = 15\,\msun$ and $P =1000$\,days.  The dotted
           curves are calculated by varying $\sigbin^2$ with factors
           of two for each subsequent curve.  The symbols with error
           bars are the observed values for these parameters from
           Table\,\ref{tab:data}.  The horizontal line is plotted to
           guide the eye.  } \label{fig:data}
\end{figure}

The linear Pearson correlation
coefficient \citep{PearsonCorrelationStatistics} for the logarithmic
values of the data presented in Fig.~\ref{fig:data} is $s =-0.71$ with
a significance level ($\siglev$) of $\sim$2$\times10^{-4}$, which
indicates that the observed trend is statistical significant
\footnote{The coefficient $s$ can have a value between $-1$ and $+1$, where $-1(+1)$
indicates a linear relation between the observed variables with
negative(positive) slope.  A value of $s = 0$ indicates a lack of
correlation.}.  The downward trend in Fig.~\ref{fig:data} suggest
that $\sigobs^2$ equals $\sigdyn^2$ plus a constant.  This is what
follows if all clusters are virialised and host a similar binary
population (equation~\ref{eq:mratio0}).

\begin{table*}
\caption{
Overview of the observational data used.  Values for the absolute
magnitudes, $\logage, \sigobs$ and $\reff$ were take from
(1)~\citet{2007A&A...461..471O}; (2)~\citet{2007MNRAS.382.1877M};
(3)~\citet{2008MNRAS.383..263L}; (4)~\citet{2006MNRAS.370..513S};
(5)~\citet{2007ApJ...663..844M};
(6)~\citet[][B06]{2006A&A...448..881B} and references therein;
(7)~\citet{2003ApJ...596..240M};
(8)~\citet[][M08]{2008A&A...489.1091M}.  The $M_V$ value of
NGC~6946-1447 was taken from the update given
by \citet{2006AJ....131.2362L}.  The values for $\mdynobs$ and
$\mphot$ were re-derived in this study, see Section~\ref{sec:data} for
details. }
\label{tab:data}      
\begin{center}
\resizebox{0.99\textwidth}{!} {
\begin{tabular}{llrrrrrrrrrr}
\hline
   Galaxy &    ID &       Ref &   $M_V$ &   $M_K$ &         $\logage$ &         $\sigobs$ &           $\reff$ &        $\mdynobs$ &          $\mphot$ & $\mdynobs/\mphot$ &    $\mphot/\reff$ \\
          &       &           &         &         &                 &        [$\kms$] &            [pc] &       [$\msun$] &       [$\msun$] &                 &      [$10^5\msun\textup{pc}^{-1}$] \\
\hline
ESO338-IG &    23 &     1 & $  -15.50$ & $      $ &   6.85$\pm$ 0.09 &   32.5$\pm$  2.5 &    5.2$\pm$  1.0 &(1.2$\pm$0.3)\,$10^7$ &(2.1$_{-0.6}^{+1.3}$)\,$10^6$ &    5.8$_{ -3.9}^{+  2.2}$ &    4.1$_{ -1.4}^{+  2.6}$ \\
  NGC1140 &    \#1 &     2 & $  -14.80$ & $      $ &   6.70$\pm$ 0.15 &   24.0$\pm$  1.0 &    8.0$\pm$  2.0 &(1.0$\pm$0.3)\,$10^7$ &(8.3$_{-0.6}^{+2.9}$)\,$10^5$ &   12.6$_{ -5.5}^{+  3.4}$ &    1.0$_{ -0.3}^{+  0.4}$ \\
  NGC1569 &     B &     3 & $  -12.85$ & $      $ &   7.30$\pm$ 0.10 &    9.6$\pm$  0.3 &    2.1$\pm$  0.5 &(4.4$\pm$1.1)\,$10^5$ &(6.8$_{-1.8}^{+0.9}$)\,$10^5$ &    0.6$_{ -0.2}^{+  0.2}$ &    3.2$_{ -1.1}^{+  0.9}$ \\
      M82 &    A1 &   4,5 & $  -14.84$ & $      $ &   6.81$\pm$ 0.03 &   13.4$\pm$  0.4 &    3.0$\pm$  0.5 &(1.2$\pm$0.2)\,$10^6$ &(1.0$_{-0.1}^{+0.1}$)\,$10^6$ &    1.2$_{ -0.2}^{+  0.3}$ &    3.4$_{ -0.7}^{+  0.7}$ \\
      M82 &  MGG9 &   6,7 & $      $ & $  -16.23$ &   6.90$\pm$ 0.15 &   15.9$\pm$  0.8 &    2.6$\pm$  0.4 &(1.5$\pm$0.3)\,$10^6$ &(8.4$_{-1.6}^{+8.7}$)\,$10^5$ &    1.8$_{ -1.8}^{+  0.5}$ &    3.2$_{ -0.8}^{+  3.4}$ \\
      M82 & MGG11 &   6,7 & $      $ & $  -15.75$ &   6.90$\pm$ 0.15 &   11.4$\pm$  0.8 &    1.2$\pm$  0.2 &(3.5$\pm$0.7)\,$10^5$ &(5.4$_{-1.0}^{+5.6}$)\,$10^5$ &    0.7$_{ -0.7}^{+  0.2}$ &    4.5$_{ -1.1}^{+  4.7}$ \\
  NGC1569 &     A &     6 & $  -14.10$ & $      $ &   7.08$\pm$ 0.20 &   15.7$\pm$  1.5 &    1.9$\pm$  0.2 &(1.1$\pm$0.2)\,$10^6$ &(1.2$_{-0.5}^{+1.0}$)\,$10^6$ &    0.9$_{ -0.8}^{+  0.4}$ &    6.1$_{ -2.6}^{+  5.1}$ \\
  NGC1705 &     1 &     6 & $  -14.00$ & $      $ &   7.08$\pm$ 0.20 &   11.4$\pm$  1.5 &    1.6$\pm$  0.2 &(4.7$\pm$1.4)\,$10^5$ &(1.1$_{-0.4}^{+0.9}$)\,$10^6$ &    0.4$_{ -0.4}^{+  0.2}$ &    6.6$_{ -2.9}^{+  5.5}$ \\
  NGC5236 &   805 &     6 & $  -12.17$ & $      $ &   7.10$\pm$ 0.20 &    8.1$\pm$  0.2 &    2.8$\pm$  0.4 &(4.2$\pm$0.6)\,$10^5$ &(2.1$_{-0.8}^{+1.6}$)\,$10^5$ &    2.0$_{ -1.6}^{+  0.9}$ &    0.7$_{ -0.3}^{+  0.6}$ \\
  NGC6946 &  1447 &     6 & $  -13.19$ & $      $ &   7.05$\pm$ 0.10 &    8.8$\pm$  1.0 &   10.2$\pm$  1.6 &(1.8$\pm$0.5)\,$10^6$ &(4.8$_{-0.4}^{+1.4}$)\,$10^5$ &    3.8$_{ -1.5}^{+  1.1}$ &    0.5$_{ -0.1}^{+  0.2}$ \\
  NGC4038 & W99-1 &     6 & $  -14.00$ & $      $ &   6.91$\pm$ 0.20 &    9.1$\pm$  0.6 &    3.6$\pm$  0.5 &(6.8$\pm$1.3)\,$10^5$ &(7.2$_{-3.2}^{+4.5}$)\,$10^5$ &    0.9$_{ -0.6}^{+  0.5}$ &    2.0$_{ -0.9}^{+  1.3}$ \\
  NGC4038 &W99-16 &     6 & $  -12.70$ & $      $ &   7.00$\pm$ 0.10 &   15.8$\pm$  1.0 &    6.0$\pm$  0.5 &(3.4$\pm$0.5)\,$10^6$ &(2.9$_{-0.9}^{+0.4}$)\,$10^5$ &   11.6$_{ -2.4}^{+  3.9}$ &    0.5$_{ -0.2}^{+  0.1}$ \\
  NGC4038 & W99-2 &     8 & $      $ & $  -17.40$ &   6.82$\pm$ 0.02 &   14.1$\pm$  1.0 &    8.0$\pm$  1.5 &(3.6$\pm$0.8)\,$10^6$ &(4.0$_{-0.4}^{+0.5}$)\,$10^6$ &    0.9$_{ -0.2}^{+  0.2}$ &    5.0$_{ -1.1}^{+  1.1}$ \\
  NGC4038 &W99-15 &     8 & $      $ & $  -15.50$ &   6.94$\pm$ 0.01 &   20.2$\pm$  1.5 &    1.4$\pm$  0.2 &(1.3$\pm$0.3)\,$10^6$ &(3.6$_{-0.3}^{+0.2}$)\,$10^5$ &    3.6$_{ -0.8}^{+  0.8}$ &    2.5$_{ -0.4}^{+  0.4}$ \\
  NGC4038 &  S1\_1 &     8 & $      $ & $  -15.70$ &   6.90$\pm$ 0.02 &   12.5$\pm$  3.0 &    3.6$\pm$  0.3 &(1.3$\pm$0.6)\,$10^6$ &(5.0$_{-0.5}^{+0.5}$)\,$10^5$ &    2.5$_{ -1.3}^{+  1.3}$ &    1.4$_{ -0.2}^{+  0.2}$ \\
  NGC4038 &  S1\_2 &     8 & $      $ & $  -15.40$ &   6.92$\pm$ 0.02 &   11.5$\pm$  2.0 &    3.6$\pm$  0.4 &(1.1$\pm$0.4)\,$10^6$ &(3.5$_{-0.2}^{+0.3}$)\,$10^5$ &    3.1$_{ -1.2}^{+  1.1}$ &    1.0$_{ -0.1}^{+  0.1}$ \\
  NGC4038 &  S1\_5 &     8 & $      $ & $  -14.80$ &   6.93$\pm$ 0.02 &   12.0$\pm$  3.0 &    0.9$\pm$  0.6 &(2.9$\pm$2.4)\,$10^5$ &(1.9$_{-0.0}^{+0.1}$)\,$10^5$ &    1.5$_{ -1.3}^{+  1.3}$ &    2.1$_{ -1.4}^{+  1.4}$ \\
  NGC4038 &2000\_1 &     8 & $      $ & $  -16.80$ &   6.93$\pm$ 0.02 &   20.0$\pm$  3.0 &    3.6$\pm$  1.0 &(3.3$\pm$1.3)\,$10^6$ &(1.2$_{-0.0}^{+0.1}$)\,$10^6$ &    2.7$_{ -1.1}^{+  1.1}$ &    3.4$_{ -0.9}^{+  1.0}$ \\
  NGC4038 &  S2\_1 &     8 & $      $ & $  -15.20$ &   6.95$\pm$ 0.01 &   11.5$\pm$  2.0 &    3.7$\pm$  0.5 &(1.1$\pm$0.4)\,$10^6$ &(3.0$_{-0.2}^{+0.2}$)\,$10^5$ &    3.8$_{ -1.4}^{+  1.4}$ &    0.8$_{ -0.1}^{+  0.1}$ \\
  NGC4038 &  S2\_2 &     8 & $      $ & $  -15.30$ &   6.95$\pm$ 0.01 &    9.5$\pm$  2.0 &    2.5$\pm$  0.5 &(5.1$\pm$2.4)\,$10^5$ &(3.2$_{-0.3}^{+0.2}$)\,$10^5$ &    1.6$_{ -0.7}^{+  0.7}$ &    1.3$_{ -0.3}^{+  0.3}$ \\
  NGC4038 &  S2\_3 &     8 & $      $ & $  -14.80$ &   6.95$\pm$ 0.01 &    7.0$\pm$  2.0 &    3.0$\pm$  1.0 &(3.3$\pm$2.2)\,$10^5$ &(2.0$_{-0.2}^{+0.1}$)\,$10^5$ &    1.6$_{ -1.1}^{+  1.1}$ &    0.7$_{ -0.2}^{+  0.2}$ \\
  NGC1487 &     1 &     8 & $      $ & $  -14.20$ &   6.92$\pm$ 0.03 &   13.7$\pm$  2.0 &    2.3$\pm$  0.5 &(9.8$\pm$3.6)\,$10^5$ &(1.1$_{-0.0}^{+0.2}$)\,$10^5$ &    8.7$_{ -3.4}^{+  3.2}$ &    0.5$_{ -0.1}^{+  0.1}$ \\
  NGC1487 &     2 &     8 & $      $ & $  -14.20$ &   6.93$\pm$ 0.02 &   11.1$\pm$  1.8 &    1.0$\pm$  0.3 &(2.8$\pm$1.2)\,$10^5$ &(1.1$_{-0.1}^{+0.1}$)\,$10^5$ &    2.5$_{ -1.2}^{+  1.1}$ &    1.1$_{ -0.3}^{+  0.4}$ \\
  NGC1487 &     3 &     8 & $      $ & $  -13.40$ &   6.93$\pm$ 0.02 &   14.3$\pm$  1.0 &    1.8$\pm$  0.3 &(8.3$\pm$1.8)\,$10^5$ &(5.3$_{-0.3}^{+0.7}$)\,$10^4$ &   15.7$_{ -4.0}^{+  3.5}$ &    0.3$_{ -0.1}^{+  0.1}$ \\
\hline
\end{tabular}
}
\end{center}
\vspace{-0.25cm}
\end{table*}

\section{Conclusions  and discussion}
\label{sec:conc}
Several studies have found from spectroscopic analyses that for many
young ($\sim$10\,Myr) star clusters the measured velocity dispersion
is too high for the mass derived from their total luminosities and
their ages. This has led several authors to conclude that these
clusters are super-virial and thus dissolving. However, the conversion
from velocity dispersion to mass (equation~\ref{eq:mdyn}) does not
consider the additional velocities of binaries. K08 considered this
effect, but concluded that binaries are only important for clusters
with low intrinsic velocity dispersion ($\sim$1\,$\kms$), i.e. lower
than the aforementioned clusters. K08 ignored the mass dependent
mass-to-light ratio of stars and the intrinsically different binary
properties of massive stars. In this study we show that taking these
aspects into account makes the contribution of binarity to the
dynamical mass estimates, $\mdynobs$, of clusters in this age range
non-negligable.

We present a simple analytical model that gives the 1-dimensional
velocity dispersion of a virialised star cluster hosting a binary
population. The model is complementary to the classical virial
relation for clusters consisting of single stars
(equation~\ref{eq:mdyn}). The result is presented as a single equation
that needs as input the (typical) binary fraction, mass ratio, primary
mass and orbital period of the binary population and the mass and
radius of the star cluster. This relation can be used to easily
estimate the effect of binaries based on different parameters for the
binary population and/or cluster.  The model presented here serves as
a starting point for more realistic approaches using binary population
synthesis models \citep[e.g][]{2009arXiv0908.1386E}. Tentative
confirmation of our results comes from the velocity dispersion of the
binary population discussed in E08: $\sim$12\,$\kms$ at an age of
10\,Myr (Eldridge 2009, priv. comm.), which is close to what we find
for the reference values discussed in Section~\ref{sec:model} (see
Fig.~\ref{fig:sig}).

For 24 clusters we derive the ratio of $\mdynobs$ over the photometric
mass, $\mphot$, and show that it decreases with increasing cluster
velocity dispersion. This is also what follows from the model and most
of the empirically determined $\mdynobs/\mphot$ ratios can be
explained by binaries using a conservative binary fraction of 25\%, a
mass ratio of $0.6$ and an orbital period of a 1000 days. When
allowing a spread in the binary parameters, almost all clusters are
within 2-standard deviation of the model results.

The fact that $\mdynobs$ and $\mphot$ generally agree for older
($\gtrsim100\,$Myr) clusters is consistent with this binary
scenario. In older clusters, we indeed expect a lower velocity
contribution of binaries. The primary star will be of a later spectral
type, thus $m_1$ is lower.  At 100\,Myr the most luminous stars are
roughly $5\,\msun$. Equation~(\ref{eq:sigbinsq}) shows that when $m_1$
is a factor of 3 lower, $\sigbin^2$ is a factor of $\sim$2 lower,
keeping all other parameters fixed. Also, typical periods are
longer.  \citet[][]{1991A&A...248..485D} find that the median period
of solar type stars is 180\,yr.  From equation~(\ref{eq:sigbinsq}) we
can see that the effect of such binaries on $\sigbin^2$ is about a
factor of $\sim$15 less than the (early type) binaries considered
here.

As mentioned in Section~\ref{sec:data}, the estimated $\mphot$
following from a comparison with SSP models, or from an IMF
extrapolation from the number of RSGs as is done in resolved clusters,
is also affected by binarity \citep{2009ApJ...696.2014D} The fraction
of stars that is removed from our sample due to this effect is
$(1-g)f$, corresponding to 45\%, giving rise to
$\mdynobs/\mphot\approx2$ for the values of Table~\ref{tab:param}.}
There is no reason, however, to expect that this would preferentially
affect clusters with low ratios $\mphot/\reff$ and it can thus
not cause the downward trend seen in Fig.~\ref{fig:data}.
 
The values of the binary parameters used in the study
(Table~\ref{tab:param}) are only indirectly based on observations
since we have to correct the period distribution found for massive
main-sequence stars to account for the reduced RSG phase of stars in
tight binaries (section~\ref{ssec:obs_bin}).  Our assumption can be
verified once the binary fraction $\frsg$ and the associated period
distribution among a statistically significant sample of resolved SGs
has been determined.  This could be done spectroscopically using a
long time base ($\sim$few~$100-1000\,\dr$).  The recently
discovered RSG clusters towards the Galactic
centre \citep{2006ApJ...643.1166F, 2007ApJ...671..781D,
2009A&A...498..109C} provide an excellent opportunity to do this. All
three have approximately the same age as the extra-galactic clusters
used here and their masses are relatively low (few times
$10^4\,\msun$) and have radii of a few pc, which according to
our model places them in the regime where binaries dominate the
measured velocity dispersion.  The ratio $\mdynobs/\mphot$ was
determined for two of them and is $\sim2$ \citep{2008ApJ...676.1016D},
lower than the extra-galactic clusters with comparable $\mphot/\reff$
(Fig.~\ref{fig:data}), but still consistent with the lower 2$\sigma$
line of our prediction. This result is very sensitive to low number
statistics since the number of RSG in these clusters is $\sim$20, so
for $\frsg=0.25$ we expect only a handful of binaries.

\citet{2009arXiv0909.3815R} present a
spectroscopic multi-epoch survey of luminous evolved stars in
Westerlund~1. This cluster is slightly younger than the clusters
considered here, thus its supergiants population is formed by more
massive stars. They find a binary fraction in excess of $40\%$ among
the 20 most luminous supergiants. Interestingly, they also find radial
velocity changes of $\sim$$15-25\,\kms$ in cool hypergiants due to
photospheric pulsations.  Macro turbulence dispersions of
$5-10\,\kms$ are also found for luminosity class II and III giants by
\citet{1986ApJ...310..277G} and \citet{2008AJ....135..892C}. This is  an additional
complication in dynamical mass determinations of young star clusters
containing massive giants.

Our results are an important ingredient in the discussion on the
importance of the early mass independent disruption, or `infant
mortality', of star clusters. The high velocity dispersions found for
the clusters discussed here have been put forward as empirical
evidence that many young ($\lesssim30\,$Myr) clusters are quickly
dissolving \citep[e.g.][M08]{2006MNRAS.373..752G}. We have provided
arguments that the alleged super-virial state can largely be explained
by orbital motions of binary stars.

Early dissolution due to gas expulsion can still exist, but it
probably occurs on much shorter time-scales ($<<10\,$Myr) than
generally assumed.  This idea is supported by the fact that the
clusters considered here have densities of $\sim$$10^3\,\msunpc$,
corresponding to an internal crossing times of the stars of roughly
$0.5\,$Myr. So these clusters have evolved for at least 20 crossing
times. The crossing time in the embedded phase is much shorter
than the crossing time at 10\,Myr due to the nonzero star formation
efficiency and the consequent expansion \citep{2008MNRAS.389..223B}.
The gas expulsion models show that clusters need about $20$ initial
crossing times to find a new virial equilibrium, or completely
dissolve into the field \citep[e.g.][]{1997MNRAS.286..669G,
2001MNRAS.323..988G, 2007MNRAS.380.1589B}.  So at 10\,Myr the
super-virial state is undetectable and the clusters discussed here are
therefore survivors of the gas expulsion, or `infant mortality',
phase.

\section*{Acknowledgement}
The authors thank John Eldridge and Selma de Mink for helpful
discussions on the evolutionary path of massive binaries and the referee, Ben Davies, for helpful suggestions. This work
was supported by NWO grant \# 639.073.803.

\bibliographystyle{mn2e}

\end{document}